\documentstyle[12 pt]{article}

\begin{document}

\textheight 9.0in
\topmargin -0.5in
\textwidth 6.5in
\oddsidemargin -0.01in

\bigskip
\begin{center}
{ \Large {\bf Quantum mechanics without spacetime\footnote {based on a talk given at the 4th International Conference on Quantum Theory and Symmetries,  Varna, Bulgaria, 15-21 August, 2005}  }}\\
{\large {\bf - A case for noncommutative geometry - }}

\bigskip

\bigskip

{{\large
{\bf T. P. Singh\footnote{e-mail address: tpsingh@tifr.res.in} }}}

\medskip

{\it Tata Institute of Fundamental Research,}\\
{\it Homi Bhabha Road, Mumbai 400 005, India.}
\vskip 0.5cm
\end{center}

\bigskip

\begin{abstract}

\noindent Quantum mechanics in its presently known formulation requires
an external classical time for its description. A classical spacetime 
manifold and a classical
spacetime metric are produced by classical matter fields. In the absence of
such classical matter fields, quantum mechanics should be formulated without
reference to a classical time. If such a new formulation exists, it follows
as a consequence that standard linear quantum mechanics is a limiting case
of an underlying non-linear quantum theory. A possible approach to the new
formulation is through the use of noncommuting spacetime coordinates in
noncommutative differential geometry. Here, the non-linear theory is
described by a non-linear Schrodinger equation which belongs to the
Doebner-Goldin class of equations, discovered some years ago. This 
mass-dependent non-linearity is significant when particle masses are
comparable to Planck mass, and negligible otherwise. Such a non-linearity
is in principle detectable through experimental tests of quantum mechanics
for mesoscopic systems, and is a valuable empirical probe of theories of
quantum gravity. We also briefly remark on the possible connection 
our approach could have with loop quantum gravity and string theory.

\end{abstract}

\newpage

\noindent {\it ``There is no doubt that quantum mechanics has seized hold
of a beautiful element of truth and that it will be a touchstone for a future
theoretical basis in that it must be deducible as a limiting case from that
basis, just as electrostatics is deducible from the Maxwell equations of
the electromagnetic field or as thermodynamics is deducible from statistical
mechanics. I do not believe that quantum mechanics will be the starting 
point in the search for this basis, just as one cannot arrive at the 
foundations of mechanics from thermodynamics or statistical mechanics.''}

\rightline{- Einstein (1936)}

\section{Why quantum mechanics without classical spacetime?}

There are several, well-known, `classical' elements in the presently known
formulation of quantum mechanics. The presence of these classical elements
in the theory makes this formulation incomplete, because such 
elements are part of a classical world. The classical world is a limiting case
of a quantum world, and a fundamental theory should not have to depend on 
its own limit, for its formulation. This suggests that a formulation more 
complete than the present one should exist, which does not refer to these 
classical elements. 

These classical elements include: (i) a prior knowledge of the canonical 
position and momenta (so that canonical commutation relations can be set up) 
and Hamiltonian, of the corresponding classical mechanical system, (ii) the 
necessity to have a concept of time [which is part of a classical spacetime 
geometry] and of a spacetime metric, so as to be able to describe quantum 
evolution, and (iii) the necessity of having a classical measuring apparatus, 
so as to be able to interpret a quantum measurement.  

These classical elements are part of the classical Universe in which we live,
which consists of classical matter fields and the classical spacetime manifold,
whose pseudo-Riemannean geometry is determined by the classical matter fields 
via the general theory of relativity. Because such a classical Universe is 
available to us today, we use its features (classical spacetime and classical 
matter) for describing a quantum system. In principle, however, such a 
classical Universe may not be available, and we should still be able to 
describe the dynamics of the quantum system.

For instance, soon after the Big Bang, there would have been a phase when 
neither the matter fields nor gravity were classical, and in this phase the 
laws of quantum mechanics could not be expressed in the form that we know
them today. One could also construct a thought experiment where a similar
situation would arise. Let us imagine a box of electrons, and imagine this to 
be the whole Universe. It is plausible to expect that the dynamics of the 
electrons should be described by quantum mechanics; however the spacetime 
metric inside the box undergoes quantum fluctuations, because the metric is 
being produced by the electrons themselves. Thus one could not write down the 
quantum dynamics of the electrons in the usual manner, because an external
classical spacetime geometry is no longer available.

One could ask if the presence of an underlying classical spacetime manifold
could be assumed, even if the overlying spacetime metric is subject to
quantum fluctuations. While this may be acceptable at an intermediate stage in
the development towards the final formulation of quantum mechanics, it cannot 
be an acceptable assumption in the final formulation of the theory. This is 
because, in the `active' picture of spacetime coordinate transformations, 
according to which coordinate transformations move points in spacetime, 
quantum fluctuations of the spacetime metric clearly destroy the point 
structure of the underlying spacetime manifold.    

It hence follows, in view of the arguments presented above, that it is
reasonable to demand that a formulation of quantum mechanics which does not 
refer to an external classical time ought to exist. In order to describe the
quantum dynamics, this new formulation will have to also incorporate an
appropriate notion of a `quantum spacetime', and hence will be intimately
related to a quantum theory of gravity. 

The new formulation should have the following two properties. Firstly, in the 
limit in which the quantum system under consideration becomes macroscopic, 
the `quantum spacetime' should become the standard classical spacetime 
described by the laws of special and general relativity; and the quantum 
dynamics should reduce to standard classical dynamics on this spacetime. 
Secondly, consider the situation in which a dominant part of the quantum 
system becomes macroscopic and classical, and a sub-dominant part remains 
microscopic and quantum. By virtue of the first property, the dominant part
should look like our classical Universe - classical matter existing in a    
classical spacetime. Seen from the viewpoint of this dominant part, the
quantum dynamics of the sub-dominant part should be the same as the standard 
quantum mechanics on an external classical spacetime.  

It is important to note that the need for such a new formulation of quantum
mechanics arises not only at the Planck energy scale. The need arises in any
situation when a background classical spacetime is not available - like
in the case of the thought experiment about the box of electrons discussed
above. This situation is not necessarily tied to the Planck scale, and the
energy scale of the electrons in question can be much below the Planck scale.
In this sense, quantum gravity, which is closely tied to the new formulation,
is not exclusively a Planck scale phenomenon.

Another motivation for developing a new formulation of quantum mechanics 
comes from the search for the correct theory of quantum gravity. Attempts
to develop a quantum theory of gravity by `quantizing' a classical theory
of gravity do face an issue of principle, because one is in this case using 
the rules of quantum theory to quantize the very spacetime geometry whose 
existence is a pre-requisite for formulating these rules. While this may
still lead to the correct quantum gravity one is nonetheless left with a 
sense of discomfort with regard to such an attempt, because it does not
seem to be the most logical way to approach the problem. In particular, as is
well-known, one is confronted with the severe unresolved `problem of time' in 
quantum gravity. It is much more desirable that the new formulation of
quantum mechanics, and the related quantum gravity, be constructed from first 
principles [without resorting to `quantization' of a classical system] and 
that standard quantum theory as well as classical general 
relativity should emerge as approximations to the new formulation, in a 
suitable limit. Eventually, only experiments can decide as to which
of the approaches to quantum gravity is the correct one.

One could object that quantum mechanics, which considers systems with finitely
many degrees of freedom, is only an approximation to quantum field theory.
Hence, once should address the issue of a new formulation also in quantum field
theory, and not just in quantum mechanics. While this would be a valid 
objection, one would naturally like to address the simpler problem first, and
only then attempt a generalization to quantum field theory.

The paper is organized as follows. In Section 2 we give an argument as to
why quantum gravity should be a non-linear theory - this argument
is independent of any specific mathematical structure one might use
to describe a quantum theory of gravity, and is hence generic.
In the third section we propose that noncommutative differential geometry
is the appropriate language for the new formulation of quantum 
mechanics. Noncommutative geometry provides a natural generalization of general
covariance to include noncommuting coordinate systems. In Section 4 we
propose a description of the quantum Minkowski spacetime using noncommuting
coordinates and  explain how standard quantum dynamics can be recovered
as an approximation to the dynamics on the noncommutative Minkowski
spacetime.

In Section 5 we develop a description for a nonlinear quantum mechanics,
to which the linear theory is an approximation. A key feature is the
introduction of an antisymmetric tensor field, which vanishes in the
classical limit, and hence explains why Riemannean geometry is a natural
approximation to noncommutative geometry, on macroscopic scales. The
non-linear Schrodinger equation we arrive at belongs to the Doebner-Goldin
class of non-linear Schrodinger equations - this aspect is discussed in
Section 6. In Section 7, we briefly emphasize the importance of experimental
tests of quantum mechanics for mesoscopic systems. 

We now show that if a formulation of quantum mechanics which does not refer
to an external classical time exists, then such a formulation is an 
approximation to an underlying non-linear quantum theory. It follows as a 
consequence that standard quantum mechanics on a background classical 
spacetime is also an approximation to a more general, non-linear quantum 
mechanics. 

\section{Why a quantum theory of gravity should be nonlinear?}

Consider the aforesaid thought experiment, wherein we consider a box of 
point particles, having masses $m_i$. The only known 
fundamental mass with which these masses can be compared is the Planck 
mass $m_{Pl}\equiv (\hbar c/G)^{1/2} \sim 10^{-5}$ grams. Since we know from
observations that microscopic masses obey quantum mechanics and macroscopic
masses obey classical mechanics, we will assume a particle behaves quantum
mechanically if its mass is much smaller than Planck mass, and classically
if its mass is much greater than Planck mass. In order to argue
that quantum gravity should be a non-linear theory we construct a series of
thought experiments where the values of the masses are made to vary, from
one experiment to the next.

We shall first consider the case where each of the masses, as well as the 
total mass of the system, is much, much smaller than Planck mass; say 
typically in the atomic mass range, or smaller. 
If we observe this box from an external classical 
spacetime, the dynamics of the particles will obey the rules of quantum
mechanics. We also assume that the total mean energy associated with
the system is much smaller than the Planck energy scale 
$E_p \sim 10^{19} GeV$. Since both Planck-mass and Planck-energy scale
inversely with the gravitational constant, the approximation we are
considering is equivalent to considering the limit 
$m_{Pl} \rightarrow\infty$, or letting $G\rightarrow 0$. It is thus
reasonable, in this approximation, to neglect the gravitational field 
produced by the particles inside the box. The reason for doing so is the 
same, for example, as to why we ignore the gravitational field of the hydrogen 
atom while studying its spectrum. What we thus have in the box is a collection 
of particles obeying quantum dynamics in an external spacetime, and the 
gravitation of these particles can be neglected.

Let us imagine now that the external classical spacetime is not there,
and that the `box' of particles is the whole universe. The arguments of the
previous section imply that there is no longer any classical spacetime
manifold available, and one should describe the quantum dynamics without
reference to it. We assume that there is a concept of `quantum spacetime'
associated with the system, and since the associated gravitational field
can be neglected we call this quantum spacetime the `quantum Minkowski
spacetime'. The classical analog of this situation is a set of particles
of small mass, whose gravitation can be neglected, existing in Minkowski
spacetime. 

The new quantum description of the dynamics of the particles in the box should 
become equivalent to standard quantum mechanics as and when a classical 
spacetime manifold becomes externally available. A classical spacetime would 
become externally available if outside the box there are classical matter 
fields which dominate the Universe.

Consider next the box universe, in the case in which, to a first order 
approximation, the values of the masses $m_i$ (as well as the total mass and 
the mean energy of the system)
in the box are no longer negligible, compared to Planck mass. 
We need to now take into account the `quantum gravitational field' produced by 
these masses, and denoted by a set of variables, say $\eta$. 
There is still no background classical spacetime manifold, but only
a background quantum Minkowski spacetime. Let us associate 
with the system a physical state $\Psi(\eta,m_{i})$. It is plausible that to 
this order of approximation the physical state is determined by a linear 
equation
\begin{equation}
\hat{O} \Psi(\eta,m_{i})=0
\label{lingrav}
\end{equation}
where the operator $\hat{O}$, defined on the background quantum Minkowski
spacetime, depends on the gravitational field variables $\eta$ only via the
linearized departure of $\eta$ from its `quantum Minkowski limit' and 
furthermore, does not have any dependence on the physical state $\Psi$.
The classical analog of this situation is a linearized description of
gravity, obtained from general relativity, when the spacetime metric is
a small departure from Minkowski spacetime, and the gravitation of the matter 
sources cannot be entirely neglected.

We now consider the case of central interest to us, where we see departure
from linear quantum theory, and argue that quantum gravity should be
non-linear. Let the masses $m_i$ in the box (as well as the total mass and 
the mean energy) be comparable to Planck mass. The behavior of the particles
will still be quantum mechanical and there is still no background classical
spacetime manifold available. Furthermore, the `quantum gravitational field'
of the system can no longer be neglected, nor is it a first order departure
from the `quantum Minkowski spacetime'. As a result, the quantum gravitational
field described by the physical state $\Psi(\eta,m_{i})$ will have
to be taken into account, and will act as a source 
for itself. This will happen recursively, so that as a consequence 
the operator $\hat{O}$ in (\ref{lingrav}) will itself depend non-linearly on 
the state $\Psi$. In this sense quantum gravity should be a non-linear theory,
because the quantum gravitational field will act as a source for itself.

We now carefully examine various aspects of this argument. There is a
well-known, useful, parallel of this argument in classical gravitation.
Starting from a linear theory of gravitation, and allowing gravity to
be a source for itself, and doing this iteratively, one concludes that
classical gravity is a non-linear theory, and one arrives at the general 
theory of relativity. The Einstein tensor of course depends non-linearly
on the spacetime metric. The above argument for a non-linear quantum gravity
is indeed analogous to the classical argument.

In this argument, the role played by the requirement for a new formulation
of quantum mechanics (which does not refer to a classical spacetime
manifold) is indirect, but crucial. We are no longer constrained by an
a priori knowledge or requirement of the standard linear quantum theory; 
instead we allow for the theory to be built from first principles, without
recourse to the aforementioned classical elements. If we stick
to the linearity constraint, then the quantization of general 
relativity naturally leads to the Wheeler-DeWitt equation - here the 
non-linearity of the gravitational field is contained, as in the classical
theory, in the three-metric appearing in the Hamiltonian constraint. However,
in our non-linearity argument above, wherein there is no classical spacetime
manifold, nor a classical spacetime metric, the only way to capture the 
information that `the quantum gravitational field acts as a source for 
itself' is via the physical state $\Psi(\eta,m_i)$ - leading to a non-linear 
quantum gravity.
 
One should contrast this situation with that for a quantized non-abelian
gauge theory. In the case of gravity, the operator $\hat{O}$ captures
information about the `evolution' of the quantized gravitational field, and
hence should depend on $\Psi(\eta,m_{i})$, because the gravitational field also
plays the role of describing spacetime structure. In the case of a non-abelian
gauge theory, the analog of the operator $\hat{O}$ again describes evolution
of the quantized gauge field, but the wave-functional $\Psi_{A}(A_i)$ 
describing the gauge field will not contribute to $\hat{O}$, because the
gauge-field does not describe spacetime structure. $\Psi_{A}(A_i)$ can, on
the other hand, be thought of as contributing non-linearly to a description
of the quantized geometry of the internal space on which the gauge field lives.

In approaches to quantum gravity wherein one quantizes a classical theory of
gravitation, using the standard rules of quantum theory, linearity is inherent,
by construction. Such a treatment could by itself yield a self-consistent
theory of quantum gravity. However, by requiring that there be a formulation
of quantum mechanics which does not refer to a spacetime manifold, one is led
to conclude that quantum gravity should be a non-linear theory. This happens
because we have introduced the notion of a quantum Minkowski spacetime 
(unavoidable when there is no classical spacetime manifold available); 
iterative corrections to this quantum Minkowski spacetime because of gravity 
bring about the non-linear dependence on the physical quantum gravitational 
state. Eventually, only experiments can decide as to whether the linear
theory is the right one, or the non-linear theory is.

The final case of the `box' Universe that should be mentioned is when all
the masses $m_i$ have values far exceeding Planck mass, in which limit
the `quantum spacetime' should reduce to standard classical spacetime
described by general relativity.

We now present an argument to show that if the quantum theory of gravity
is non-linear, the equation of motion which describes the quantum dynamics
of the particles in the box is also non-linear, and provides a non-linear
generalization of the Schrodinger equation. We do this, like above, by
considering thought experiments with different values for masses of the
particles.

An analogy with classical general relativity will be helpful. On a classical 
background Minkowski spacetime, the equations of motion of particles are those
provided by special relativity. On a curved classical background, the motion 
of a test particle is geodesic. If the particle is not a test particle but 
contributes to the gravity of the spacetime, the motion of the particle, and 
the gravitational field of the spacetime, have to be determined 
self-consistently, and this implies corrections to the geodesic equation of 
motion. A similar argument for the case of quantum mechanics without a 
classical spacetime manifold implies that there should be non-linear 
corrections to the Schrodinger equation, as we demonstrate below.

Consider the dynamical equation which describes the motion of a 
particle $m_1$ in the box universe, in the absence of a classical spacetime 
manifold. We assume that there are also present other particles, and in 
general $m_1$  and all the other particles together determine the 
`quantum gravitational field' of this `quantum spacetime'.

The simplest situation is the one in which the masses of all the particles are
much smaller than Planck mass, so that the spacetime is a quantum version of
Minkowski spacetime, and there is no gravity. In this case, the equation of
motion of $m_1$ will be the `classical background independent' analog of the
flat spacetime Klein-Gordon equation.
 
Keeping the mass $m_1$ small, increase the mass of the other particles so
that $m_1$ becomes a test particle, while the `quantum gravitational field'
of the other particles obeys the linear equation (\ref{lingrav}).
In this case, the equation of motion of $m_1$, which is a test particle,
will be the `classical background independent' analog of the curved spacetime 
Klein-Gordon equation, where the `quantum gravitational field' depends 
linearly on the source.

Still keeping the mass $m_1$ a test particle, 
increase the mass of the other particles further, to the Planck mass range,
so that the `quantum gravitational field'
of the other particles now obeys a non-linear equation, where
the quantum gravitational field depends non-linearly on the quantum state
of these particles. In this case, the equation of motion of $m_1$ will be 
the `classical background independent' analog of the curved spacetime 
Klein-Gordon equation, where the `quantum gravitational field' now depends 
non-linearly on the quantum state of the source, the source being all the 
particles other than $m_1$.

Now increase the mass $m_1$ also, to the Planck mass range, so that it is no
longer a test particle. The equation of motion of $m_1$ thus now depends
non-linearly on the quantum gravitational state of the system, where system
now includes the mass $m_1$ as well. A special case would be one where we 
remove all particles, except $m_1$, from the system. The equation of motion for
$m_1$ would then depend non-linearly on its own quantum state. This is what 
gives rise to non-linear quantum mechanics, which when seen from a classical 
spacetime manifold (as and when the latter becomes available) would be a 
non-linear Schrodinger equation. The possibility of non-linear modifications
to the Schrodinger equation arises here because one has not a priori imposed
the constraint that quantum mechanics is a linear theory on a background
classical manifold. Instead, the quantum mechanics is being inferred from
first principles.

It can be inferred that the introduction of non-linearity in quantum gravity 
and quantum mechanics is related to the introduction of the (Planck) mass 
(equivalently energy) scale. When the masses in question are much smaller than 
Planck scale, the situation is equivalent to sending Planck mass to infinity, 
and also, the theory is then linear. There is thus a direct connection between 
non-linearity in quantum mechanics, and the presence of a fundamental energy
scale in the theory. 

Although the following is not a consequence of the above discussion, nor needed
for further discussion, we would like to suggest a structure for the quantum
Minkowski spacetime. This spacetime does not have any information about the
masses present, does not have gravity, nor is it classical. Hence a plausible 
structure for it is that it is a grid made of cells of Planck length size. A 
coarse graining on scales larger than Planck length gives makes it appear
as the classical Minkowski spacetime. 

If this picture is correct, it implies a nice parallel between classical and
quantum gravity. Classical gravity, as described by general relativity,
imposes the gravitational field, produced by classical matter, on 
Minkowski spacetime. Quantum gravity, in our picture, imposes a `quantum
gravitational field', produced by quantum matter, on the quantum Minkowski
spacetime possessing a Planck size grid. Coarse graining of this grid, along
with the classical limit for matter, should reduce quantum gravity to
classical general relativity. Thus, for us, quantum gravity is not the 
`quantization' of classical general relativity, but a theory completely
parallel to the classical theory, and built on the quantum Minkowski spacetime,
instead of the classical Minkowski spacetime. Both the theories of gravity, 
classical as well as quantum, are non-linear.

Lastly in this section, we would like to comment on the significance of
Planck mass ($\sim 10^{-5}$ grams), versus Planck energy ($\sim 10^{19}$ Gev).
Because of the mass-energy equivalence, they are both of course the same
concept. In the literature however, an impression is sometimes conveyed
that Planck energy is a fundamental scale, whereas Planck mass is not. 
This viewpoint is partly unavoidable because, as masses go, Planck mass is 
a rather large mass - it
is comparable to the mass of a small grain of dust, pretty much in the 
classical range. Nonetheless, if new physics arises at the Planck energy 
scale, it should definitely arise also at the Planck mass scale. 
Considering that the value of Planck mass is embarrassingly large, what is 
likely is that if new physics does arise, it comes about at a fundamental 
scale $m_c$ which is a few orders below the Planck mass scale (and 
equivalently a few orders below the Planck energy scale). Why $m_c$ should be
not the same as Planck mass, but smaller, is not clear, and we will continue
to refer to the fundamental mass scale as Planck mass $m_{Pl}$, although what
we really have in mind is the mass scale $m_c$. 

Continuing on this issue, it is said that quantum gravity effects can be probed
by accelerating an elementary particle (whose rest mass is much smaller than
Planck mass) to Planck energies, so that it can probe length scales as
small as Planck length. While this is of course true, it would certainly 
also be possible to probe quantum gravity effects with an elementary particle
(if such a particle existed) whose rest mass becomes comparable to Planck 
mass, so that its Compton wavelength becomes comparable to Planck length.
Known objects with mass in the Planck mass range are much, much larger than
Planck length, in physical size, because of non-gravitational effects (the
nuclear and the electromagnetic force). This rules them out as probes of
physics on the Planck length scale. (It is also interesting to note that
Planck mass is perhaps the mass scale where classicality sets in, because for
objects with a mass larger than Planck mass the Schwarzschild radius becomes
larger than the Compton wavelength, so that the quantum nature of the object is
hidden behind the Schwarzschild radius.)

However, one cannot a priori rule out the possibility that Planck mass
objects can serve as probes of non-linear corrections that might arise in 
quantum mechanics, near the Planck mass scale. This is especially significant 
in light of the fact that quantum mechanics has indeed not been 
experimentally tested for isolated mesoscopic systems whose mass lies in the 
intermediate range (say an object having a mass of a billionth of a gram) - in 
between the mass of an atomic system and the mass of a macroscopic object.     
The non-linear corrections could show up, for instance, as corrections to
the energy levels of a system, which could happen independent of the fact 
that a realistic Planck mass object may not probe physics at the Planck 
length scale.

The arguments and conclusions presented in these two sections are generic,
and do not depend on the specific mathematical structure used to develop
a quantum theory of gravity. The only key assumptions made are that there
should be a new formulation of quantum mechanics which does not refer to
a classical spacetime manifold, and that the quantum gravitational field acts
as a source for itself, like the classical gravitational field does, in the
classical theory. In the next section we outline the proposal that the 
new formulation should be developed using the language of noncommutative 
differential geometry.

\section{A formulation based on noncommutative differential geometry}

Noncommutative differential geometry is an abstract, but in retrospect
a rather natural, extension of Riemannean geometry, that includes the latter
as a special case. (A short but highly readable account, on which the present 
brief discussion is based, is by Martinetti \cite{mar}). The starting point of 
the development is the famous theorem of Gelfand, Naimark and Segal which 
shows that the topological properties of a compact space $X$ can be described 
completely in algebraic terms, i.e. in terms of the algebra of complex 
functions on $X$. Conversely, given an algebra of functions ${\cal A}$ (with a 
few basic additional properties) one can build a compact topological space 
$X$, for which ${\cal A}$ can be interpreted as the algebra of functions. One 
could also generalize the algebra ${\cal A}$ to a noncommutative algebra, and 
build from it a noncommutative space $Y$, for which ${\cal A}$ can be
interpreted as the algebra of functions.  

One could next ask about the coding of the differential structure of a manifold
in algebraic terms. Connes observed that, given a Riemannean spin manifold, 
its geometrical information can be recovered from the so called spectral 
triple, which consists of an algebra ${\cal A}$, a Hilbert space ${\cal H}$ and
a self-adjoint operator ${\cal D}$. Thus one has a map from a Riemannean spin
geometry to an algebra, which is commutative. Conversely, given a 
commutative algebra ${\cal A}$ as part of a spectral triple, one can build
from it a Riemannean spin manifold.

The central point now is that the features used in mapping from the algebra
to geometry do not depend on the commutativity of the algebra. One can make
the algebra of functions noncommutative - the geometry to which the algebra
then gets mapped is called noncommutative differential geometry. Riemannean 
geometry becomes a special of Connes' noncommutative geometry. Also, a 
concept of distance and metric can be developed for the noncommutative
space, given the operator ${\cal D}$.     

Given the map between a differentiable manifold and the corresponding 
commutative algebra, diffeomorphisms on the manifold can be mapped to the 
automorphisms of the algebra.  When the algebra is noncommutative, the 
automorphisms of the algebra represent corresponding diffeomorphisms on the 
noncommutative space. Thus if the noncommutative space possesses a symmetry 
representing invariance under diffeomorphisms, this symmetry can be expressed 
as an invariance under the algebra automorphisms, and could be called 
automorphism invariance. Of special interest are the inner automorphisms,
which are implemented by unitary elements of the algebra - these form a group,
which is trivial in the commutative case. The inner automorphisms can hence
be interpreted as the noncommutative part of the diffeomorphisms on the
noncommutative space.

Our proposal is that noncommutative differential geometry is an appropriate
framework for a formulation of quantum mechanics which does not refer to a
classical (i.e. Riemannean) spacetime manifold. We will motivate this approach,
and describe its present status, in the following sections. Our present
understanding, though well-motivated on physical grounds, is still partly
heuristic, and does not yet make contact with the rigorous formulation of 
noncommutative geometry, for instance with regard to construction of the 
spectral triple, and relating the noncommutative metric introduced below to 
the one formally defined in noncommutative geometry. Nonetheless the ideas 
developed below hold out the promise that the formal connections of these 
ideas with noncommutative geometry will eventually get developed.

We begin by suggesting the notion of a noncommuting coordinate system, which
is to be thought of as `covering' a noncommutative manifold, and wherein the
commutation relations between the coordinates are to be introduced on
physical grounds. The new formulation of quantum dynamics is to be given
in such a coordinate system, and is to be invariant under transformation 
(an automorphism) from the given coordinate system to different noncommuting 
coordinate systems. This is a generalization of general covariance to the 
noncommutative case, and one is proposing that a physical theory should be 
invariant under transformations of noncommuting coordinates. 

By drawing analogies with special and general relativity, we will argue that 
the new formulation of quantum mechanics is essentially equivalent to a 
generalization of relativity to a noncommutative spacetime. That is, when one 
makes a transition from a commutative spacetime to a noncommutative spacetime, 
one simultaneously also makes a transition from classical dynamics to quantum 
dynamics. The quantum Minkowski spacetime is arrived at by introducing 
noncommutativity on a Minkowski spacetime, and quantum gravity is arrived at 
by introducing noncommutativity on a curved classical spacetime. 

This formulation on a noncommutative spacetime satisfies the two properties
mentioned in the introduction. When the system being described becomes
macroscopic, the dynamics coincides with classical dynamics. Also, if
the quantum dynamics (expressed in terms of noncommuting coordinates) 
is viewed from an external, classical spacetime (described by ordinary
coordinates) it looks the same as standard quantum dynamics. 

It may appear surprising as to how spacetime noncommutativity, for which
the natural scale is the Planck length scale, can give rise to the standard 
quantum dynamics, which becomes relevant already on length scales much larger 
than Planck length. The answer, as we will see, lies in noting that in addition
to spacetime noncommutativity, one also introduces momentum space 
noncommutativity. The noncommutativity between coordinates (which
occurs at a very small length scale) together with noncommutativity between 
momentum components (which occurs at a very high energy scale), gives rise to 
a dynamics which looks like the standard quantum dynamics, when seen from our 
classical spacetime.

One could also ask as to why noncommutative differential geometry is the
right arena for a mathematical formulation of the physical ideas described
in the previous two sections. One could of course not give a proof for this,
but only a motivation, and eventually the predictions of the new formulation
have to be tested, and confirmed or ruled out, by laboratory experiments. The 
motivation is that noncommutative geometry is a natural extension of 
Riemannean geometry, and incorporates the generalization of diffeomorphisms 
to include transformations of noncommuting coordinates, and automorphism 
invariance is a plausible physical symmetry. It is then reasonable to ask 
what new physics could possibly relate to automorphism invariance. On the 
other hand, while noncommutativity is intrinsic to quantum mechanics, this 
noncommutativity is introduced in an ad hoc manner, building on a pre-existing
classical theory. One would like to explore whether this ad hoc element can 
be avoided by relating the feature of noncommutativity inherent in 
noncommutative geometry to the noncommutativity that is central in quantum 
mechanics. And if one wants to avoid dependence on a classical, commuting 
spacetime manifold, and yet have a concept of `quantum spacetime' from 
which ordinary spacetime emerges, a noncommutative manifold appears to be the 
most likely starting point.      

We should also compare our approach with other investigations of applications 
of noncommutative geometry to physics. While these other studies undoubtedly
explore some fascinating avenues, our approach differs from them in that we
are proposing noncommutative differential geometry as the appropriate
mathematical language for describing quantum dynamics. Amongst other 
investigations, a prominent line of research, inspired in part by string 
theory, is quantum field theory on a noncommutative spacetime. Here, 
spacetime noncommutativity (usually only space-space noncommutativity) is
assumed at the outset, as resulting from an underlying quantum structure of
spacetime. One then investigates how this noncommutativity affects 
quantization of fields, with regards to issues like  renormalizability, and 
avoidance of divergences \cite{gra}. Another important line of development is 
concerned with investigating consequences of noncommutativity for the 
structure of spacetime \cite{fre}. Spacetime noncommutativity appears to
also play an important role in studies of Doubly Special Relativity
\cite{kow} and the possibility of violation of Lorentz symmetry \cite{ame}.
Investigation of noncommutative gravity at the classical level has
also received serious attention \cite{cha}.

While it is perhaps fair to say that at present a clear picture of the
relation between noncommutative geometry and quantum gravity is not
available, we would like to mention two important ideas which in our view could
likely become center-stage in the future. The first is the possibility that
noncommutative geometry could in a natural way provide an explanation for
the origin of time, something for which it is difficult to find an explanation
in the standard approaches to quantum gravity. A key property of von Neumann
algebras is given by the Tomita-Takesaki theorem, which defines a one-parameter
group of automorphisms of the algebra, and hence allows for a time-flow - this
is the `thermal time hypothesis' \cite{con}. An alternative explanation for
the spontaneous generation of time has recently been put forward by Majid
\cite{maj} who shows this to be a natural consequence of properties of the
noncommutative differential calculus, when one starts with space-space
noncommutativity. 

The second key idea is that by taking the product of a manifold with a 
suitable finite dimensional geometry, the standard model of particle physics
can be described in the language of noncommutative geometry \cite {chc}.

The application of noncommutative geometry to physical situations is 
generally segregated from quantum theory. Often, spacetime (or space-space)
noncommutativity is introduced first, and quantization of the system is 
undertaken as a separate step. We now try to make the case that it may be
fruitful to allow spacetime noncommutativity and quantum mechanics to go
hand in hand, the latter being a consequence of the former.

\section{A possible description of quantum Minkowski spacetime}

As mentioned above, a quantum Minkowski spacetime will arise if no external
classical spacetime is available, and if the masses of all the particles in 
the box Universe (as well as the total mass and energy scale of the box) are 
much smaller than Planck mass. In order to motivate our construction of the 
quantum Minkowski spacetime using noncommuting coordinates, we first briefly 
recall the relativistic Schrodinger equation for a free particle in quantum 
mechanics, and for simplicity of presentation we assume only one space
dimension. Subsequently we will generalize to the case of many particles, and 
also to 3+1 spacetime.  We chose to consider the relativistic case, as
opposed to the non-relativistic one, only because the available space-time
symmetry makes the analysis more transparent. Later, we will consider
also the non-relativistic limit.

The relativistic Schrodinger equation for a particle of mass
$m$ in 2-d spacetime
\begin{equation}-\hbar^{2}
\left({\partial^{2}\over\partial t^{2}}-{\partial^{2}\over\partial x^{2}}\right)\psi=m^{2}\psi\label{kg}
\end{equation}
can be rewritten, after the substitution $\psi=e^{iS/\hbar}$, as
\begin{equation}
\left({\partial{S}\over \partial t}\right)^{2}-\left({\partial{S}\over \partial x}\right)^{2}
-i\hbar\left({\partial^{2}S\over\partial t^{2}}-{\partial^{2}S\over\partial x^{2}}\right)=m^{2}\label{hjc}.
\end{equation}

The object $S(t,x)$ introduced above, which we will call the complex action,
will play a central role for us. In the classical limit, it will become real 
and coincide with the classical action. We would like to suggest that
quantum dynamics could alternately be described in terms of this complex 
action, via Equation (\ref{hjc}), which should be thought of as the quantum
Hamilton-Jacobi equation. The quantum corrections of course come from the
$\hbar$-dependent terms, which correct the classical Hamilton-Jacobi equation.
In terms of the complex action $S(t,x)$ we define the generalized momenta 
\begin{equation}
p^{t}=-{\partial S\over \partial t}, \qquad p^{x} = 
{\partial S \over \partial x}
\label{pmoo}
\end{equation}
in terms of which Eqn. (\ref{hjc}) reads
\begin{equation}
p^{\mu}p_{\mu} + i\hbar {\partial p^{\mu}\over \partial x^{\mu}} = m^{2}.
\label{hjp}
\end{equation}
where the index $\mu$ takes the values $1$ and $2$. The momenta defined above
as gradients of the complex action have an obvious parallel with classical
dynamics. In terms of the complex action $S(t,x)$ we have a uniform
description of classical and quantum dynamics. Below, we will make the
case that Eqn. (\ref{hjc}) can be derived from the fundamental assumptions
of an underlying noncommutative theory.

In standard quantum mechanics the origin of the $\hbar$-dependent term in
(\ref{hjp}) of course lies in the position-momentum commutation relation, and
this term provides a correction to the classical  
energy-momentum relation $E^{2}-p^{2}=m^{2}$. This, along with the
consideration that (\ref{hjp}) should be written in a more symmetric
manner so that spatial gradients do not appear, suggests that from the
viewpoint of dynamics on a noncommutative spacetime, these correction terms
maybe expressible in terms of noncommutativity of momenta.   

Following this lead, we now suggest a model for the
dynamics of the `quantum Minkowski spacetime' by assuming that there exist
two noncommuting coordinates $\hat{x}$ and $\hat{t}$, and that the quantum
mechanical particle lives in this noncommutative spacetime. 
We ascribe to the particle a `generalized momentum' $\hat{p}$,
having two components $\hat{p}^{t}$ and $\hat{p}^{x}$, which do not
commute with each other. The noncommutativity of these momentum components
is assumed to be a consequence of the noncommutativity of the coordinates,
as the momenta are defined to be the partial derivatives of a complex action
$S(\hat{x},\hat{t})$, with respect to the corresponding noncommuting 
coordinates. We will propose a dynamics for the particle, analogous to 
classical special relativity, in these noncommuting coordinates. We will then
argue that seen from an external classical spacetime (as and when the latter
exists) this noncommutative dynamics looks the same as standard quantum
mechanics as described by Equations (\ref{kg}) and (\ref{hjc}).

On the quantum Minkowski spacetime we introduce the following noncommutative 
flat metric 
\begin{eqnarray}
\label{ncfm}  
\hat{\eta}_{\mu\nu} = \left(\begin{array}{cc}
                      1 & 1 \\
                      -1 & -1 \end{array} \right)
\end{eqnarray}
and assume that there exists a corresponding noncommutative line-element
\begin{equation}
ds^{2}=\hat{\eta}_{\mu\nu}d\hat{x}^{\mu}d\hat{x}^{\nu}=
d\hat{t}^{2}-d\hat{x}^{2}
+d\hat{t}d\hat{x}-d\hat{x}d\hat{t}.
\label{lin}
\end{equation}
In analogy with special relativity, the metric and the line-element are
assumed to be invariant under a class of coordinate transformations which
generalize Lorentz transformations. Thus this line-element is left invariant
by the transformation
\begin{equation}
\hat{x}' =  \gamma (\hat{x} - \beta\hat{t}), \qquad
\hat{t}' = \gamma (\hat{t} - \beta \hat{x})
\label{lor}
\end{equation}
where $\gamma=(1-\beta^{2})^{-1/2}$. In the commutative case, $\beta$ has the
interpretation of velocity: $\beta=v/c$. On a noncommutative space, $\beta$
could be thought of as defining a `rotation' in the noncommutative space,
by an angle $\theta$ such that $\beta=\tanh\theta$. 

The above metric, which is obtained by adding an antisymmetric component to
the standard Minkowski metric, is assumed to generalize the Minkowski metric
to the noncommutative case. It is non-Hermitean, and has zero determinant - 
below we consider also a Hermitean modification of this metric. Our discussion
here, though, is not obstructed by the fact that the metric is not Hermitean.

In order to motivate a noncommutative dynamics, we note that, from 
(\ref{lin}), one could heuristically define a velocity  $\hat{u}^{i}=
d\hat{x}^{i}/ds$,
which satisfies the relation
\begin{equation}
1=\hat{\eta}_{\mu\nu}\frac{d\hat{x}^{\mu}}{ds}\frac{d\hat{x}^{\nu}}{ds}=
(\hat{u}^{t})^{2}-(\hat{u}^{x})^{2} + 
\hat{u}^{t}\hat{u}^{x} - \hat{u}^{x}\hat{u}^{t}.  
\label{vel}
\end{equation} 
This suggests a definition of the generalized momentum as 
$\hat{p}^{i}=m\hat{u}^{i}$, 
which hence satisfies 
\begin{equation}
\hat{p}^{\mu}\hat{p}_{\mu} = m^{2}
\label{nchj}
\end{equation}
Here, $\hat{p}_{\mu}=\hat{\eta}_{\mu\nu}\hat{p}^{\mu}$ is well-defined. Written
explicitly, this equation becomes
\begin{equation}
(\hat{p}^{t})^{2}-(\hat{p}^{x})^{2} + 
\hat{p}^{t}\hat{p}^{x} - \hat{p}^{x}\hat{p}^{t}  = m^{2}
\label{nce}
\end{equation} 
Eqn. (\ref{nchj}) appears an interesting and plausible proposal for the 
dynamics, because it generalizes the corresponding special relativistic 
equation to the noncommutative case. The noncommutative Hamilton-Jacobi
equation is constructed from (\ref{nce}) by defining the momentum as gradient
of the complex action function. We are proposing here that in the absence of
a classical spacetime manifold, quantum dynamics should be described using this
Hamilton-Jacobi equation - this is the noncommutative analog of the quantum
Hamilton-Jacobi equation (\ref{hjc}).

Let us consider now that an external classical Universe becomes available 
(in the next section we will discuss how a  classical Universe could arise
from the quantum dynamics). Seen from this classical Universe the quantum
dynamics should be as described by Eqn. (\ref{hjc}). Thus, we need to justify
the following correspondence rule, when the noncommutative Hamilton-Jacobi
equation is examined from our standard spacetime point of view:  
\begin{equation}
(\hat{p}^{t})^{2}-(\hat{p}^{x})^{2} + 
\hat{p}^{t}\hat{p}^{x} - \hat{p}^{x}\hat{p}^{t}  = ({p}^{t})^{2}-({p}^{x})^{2}
 + i\hbar {\partial p^{\mu}\over \partial x^{\mu}}
\label{nceq}
\end{equation}
Here, $p$ is the `generalized momentum' of the particle as seen from the 
commuting coordinate system, and is related to the complex action by 
Eqn. (\ref{pmoo}). This equality should be understood as an equality between 
the two equivalent equations of motion for the complex action function 
$S$ - one written in the noncommuting coordinate system, and the other 
written in the standard commuting coordinate system.

The idea here is that by using the Minkowski metric of ordinary spacetime 
one does not correctly measure the length of the `momentum' vector, because 
the noncommuting off-diagonal part is missed out. The last, 
$\hbar$ dependent term in (\ref{nceq}) provides the correction - the origin
of this term's relation to the commutator 
$\hat{p}^{t}\hat{p}^{x} - \hat{p}^{x}\hat{p}^{t}$ can be understood as 
follows. 

Let us write this commutator by scaling all momenta with respect to Planck
momenta: define $\hat{P}^{\mu}=\hat{p}^{\mu}/P_{pl}$. Also, all lengths are 
scaled with respect to Planck length: $\hat{X}^{\mu}=\hat{x}^{\mu}/L_{pl}$. 
Let the components of $X^{\mu}$ be denoted as $(\hat{T}, \hat{X})$. The
commutator $\hat{P}^{T}\hat{P}^{X} - \hat{P}^{X}\hat{P}^{T}$ represents the 
`non-closing' of the basic quadrilateral when one compares (i) the operator 
obtained by moving first along $\hat{X}$ and then along $\hat{T}$, with 
(ii) the operator obtained by moving first along $\hat{T}$ and then 
along $\hat{X}$. When seen from a commuting coordinate system, this
deficit (i.e. non-closing) can be interpreted as a result of moving to a
neighboring point, and in the infinitesimal limit the deficit will be the
sum of the momentum gradients in the various directions. The deficit is thus 
given by $i{\partial P^{\mu}/ \partial X^{\mu}}=i(L_{Pl}/P_{Pl}) {\partial p^{\mu}/ \partial x^{\mu}}$. This gives that $\hat{p}^{t}\hat{p}^{x} - \hat{p}^{x}
\hat{p}^{t}=  i\hbar {\partial p^{\mu}/ \partial x^{\mu}}$.
Hence, since the relation (\ref{nceq}) holds, 
there is equivalence between the noncommutative description 
(\ref{nchj}) and standard quantum dynamics given by (\ref{hjp}).

We have not yet said anything about the all-important commutation relations
on the non-commutative spacetime. The following is a plausible, natural 
structure:
\begin{equation}
[\hat{t},\hat{x}]=iL_{Pl}^{2}, \qquad [\hat{p}^{t}, \hat{p}^{x}]=iP_{Pl}^{2}.
\label{commu}
\end{equation}
These relations look plausible because in the quantum Minkowski spacetime
gravity is ignorable; so the commutation relations should not carry any
information about the matter content of the spacetime, and hence they should
not depend on the mass $m$. It is generally not considered likely that
spacetime noncommutativity on the Planck length scale could account for quantum
effects on the much larger atomic scales. However, here we note that we have
also introduced noncommutativity on the Planck momentum scale, which of
course is much larger than the momenta encountered on atomic scales. We could
picture that on the noncommutative 4-d `phase space' corresponding to the
two noncommuting coordinates $\hat{x},\hat{t}$ there is a natural grid 
(which is a consequence of noncommutativity) which when projected on the
$\hat{x},\hat{p}^{x}$ plane, has an area $L_{Pl}\times P_{PL}=\hbar$. When we
choose to examine this noncommutative quantum dynamics from our ordinary 
spacetime, we take the limit $L_{Pl}\rightarrow 0, P_{Pl} \rightarrow \infty$,
while keeping their product (the area of the fundamental phase space cell)
constant at $\hbar$. From the viewpoint of our classical spacetime, quantum 
dynamics is then recovered by imposing the commutator $[q,p]=i\hbar$ which 
preserves the granular structure of the phase space in quantum mechanics. 

This completes our construction for the quantum dynamics of a particle
on a 2-d noncommutative spacetime. It remains to be seen whether this
dynamics can be obtained from an action principle. In the next section we 
will see how the off-diagonal part of the noncommutative metric (\ref{ncfm}) 
introduced above gets suppressed when the particle has a mass much larger 
than Planck mass, so that this construction can no longer be distinguished 
from classical dynamics.

It is straightforward to generalize this construction to 4-d spacetime. The
noncommutative metric is constructed in analogy with (\ref{ncfm}) - the
diagonal part is $(1,-1,-1,-1)$; the off-diagonal entries are $1$ and
$-1$ - those above the diagonal are $1$ and those below the diagonal are 
$-1$ - so that the off-diagonal part is antisymmetric. The energy-momentum
relation is the same as in Eqn. (\ref{nchj}) above, with the index $\mu$
running from one to four. In Eqn. (\ref{nceq}) the net off-diagonal 
contribution coming from the commutators is to be equated to the gradient of
the complex action on the right hand side, the reasoning for doing so being
the same as in the 2-d case. Furthermore, commutation relations analogous  
to those in (\ref{commu}) are assumed to hold for each coordinate pair and
each momentum pair. However an important change which comes about is that the
4-d line-element analogous to (\ref{lin}) is not invariant the Lorentz
transformation (\ref{lor}), suggesting that there is a new class of 
transformations which leave this line-element invariant, and which reduce to
Lorentz transformations in the limit of a commutative spacetime. (In fact,
the presence of a second fundamental scale, namely Planck length, suggests
the possibility of a connection with Doubly Special relativity).

The multi-particle case can be described in analogy with the way it is dealt
with in special relativity. We will continue to assume that the generalized
momentum of a particle is defined as the gradient of the complex action with 
respect to the coordinate of the corresponding particle, and that the 
energy-momentum relation (\ref{nchj}) holds separately for each of the 
particles. The total momentum is the sum of the momenta of the individual 
particles, and the quantum Hamilton-Jacobi equation is constructed by squaring 
the total momentum and using (\ref{nchj}) for each of the particle. In this 
regard, the treatment is no different from that of the multi-particle case
in special relativity.

Lastly in this section we comment on the possibility of using a Hermitean 
metric
\begin{eqnarray}
\label{nchm}  
\hat{\eta}_{\mu\nu} = \left(\begin{array}{cc}
                      1 & i \\
                      -i & -1 \end{array} \right)
\end{eqnarray}
instead of the metric (\ref{nchm}). By following the construction given above,
it is easy to see that the same dynamics can alternately be described with
this metric. What is not clear though is the implication of the fact that this
metric can be diagonalized, via a complex transformation, to Minkowski-like
coordinates. In the following, we will work with a generalization of the 
metric (\ref{ncfm}) - which will have a nice interpretation that the 
off-diagonal components become negligible in the macroscopic limit, 
so that the metric reduces to the standard metric on a curved space.

In summary, we have tried to make the case that dynamics in noncommutative 
special relativity is the same as quantum dynamics, if one assumes the
above form for the noncommutative metric, along with the proposed
commutation relations. Our discussion in this section has been largely 
heuristic, and further work needs to be done to make rigorous contact with 
noncommutative differential geometry. One could also criticize the 
construction by noting that there is no suggestion here for experimentally 
verifying whether the noncommutative description is indeed the right one. We 
will argue now that an experimental prediction can be made, by considering the 
quantum dynamics of a mass whose value approaches the Planck mass.

\section{A non-linear Schrodinger equation} 

In Section 2 we argued that when the particle masses become comparable to
Planck mass, the quantum equation of motion for the particles should become 
non-linear. We now present an approximate construction for such a non-linear 
equation, based on the dynamics for the `quantum Minkowski spacetime' proposed 
above. Since the quantum Minkowski spacetime and its metric were obtained as
a noncommutative generalization of special relativity one can expect that
there will also be a generalization of the quantum Minkowski spacetime to
a `quantum curved spacetime' which will be equivalent to a noncommutative 
generalization of general relativity. This will be the origin of non-linearity;
and since standard quantum dynamics corresponds to the dynamics on the quantum
Minkowski spacetime, one can expect that the dynamics on the quantum curved
spacetime will be a non-linear generalization of standard quantum 
dynamics. 

We once again start by considering the case of a single particle in 2-d 
spacetime. The starting point for our discussion will be Eqn. (\ref{nchj}) 
above - we will assume that a natural generalization of this equation
describes quantum dynamics when the particle mass $m$ is comparable to the 
Planck mass $m_{Pl}$, and the effects of the particle's own gravity
cannot be ignored. In this case, we no longer expect the metric $\eta $ to 
have the `flat' form given in (\ref{ncfm}) above. The symmetric components of 
the metric are of course expected to start depending on $m$ (that is gravity)
and in general the antisymmetric components can also be expected to depend on 
$m$ - so long as the antisymmetric components are non-zero, we can say that
quantum effects are present. As $m$ goes to infinity, the antisymmetric
component should go to zero - since in that limit we should recover
classical mechanics. In fact the antisymmetric part (we will call it $\theta
_{\mu \nu} $) should already start becoming ignorable when $m$ 
exceeds $m_{Pl}$. It is interesting to note that the symmetric part should 
grow with $m$, while the antisymmetric part should fall with increasing $m$. 
There probably is some deep reason why this is so. 

If $\hat{h}_{\mu\nu}$ is the noncommutative metric which generalizes 
$\hat{\eta}_{\mu\nu}$ we can write it as
\begin{eqnarray}
\label{nccm}  
\hat{h}_{\mu\nu} = \left(\begin{array}{cc}
                      \hat{g}_{tt} & \hat{\theta} \\
                      -\hat{\theta} & -\hat{g}_{xx} \end{array} \right)
\end{eqnarray}
There is a corresponding generalization of the line-element,
\begin{equation}
ds^{2}=\hat{h}_{\mu\nu}d\hat{x}^{\mu}d\hat{x}^{\nu}=
\hat{g}_{tt}d\hat{t}^{2}-\hat{g}_{xx}d\hat{x}^{2}
+\hat\theta[d\hat{t}d\hat{x}-d\hat{x}d\hat{t}].
\label{linc}
\end{equation}
while the velocity and momentum are defined as in the previous section. The 
energy-momentum relation will now be
\begin{equation}
\hat{h}_{\mu\nu}\hat{p}^{\mu}\hat{p}^{\nu}=m^{2} 
\end{equation}
and instead of Eqn. (\ref{nceq}), we will have the dynamical equation 
\begin{equation}
\label{nceq2}\hat g_{tt}(\hat p^t)^2-\hat g_{xx}(\hat p^x)^2+\hat \theta
\left( \hat p^t\hat p^x-\hat p^x\hat p^t\right) =m^2
\end{equation}
and the correspondence rule
\begin{equation}
\label{corr}\hat g_{tt}(\hat p^t)^2-\hat g_{xx}(\hat p^x)^2+\hat \theta
\left( \hat p^t\hat p^x-\hat p^x\hat p^t\right)=g_{tt}({p}^t)^2-g_{xx}({p}%
^x)^2+i\hbar \theta {\frac{\partial p^\mu }{\partial x^\mu }}
\end{equation}
and as seen from an external classical spacetime the new dynamical equation  
\begin{equation}
\label{nceq3}g_{tt}({p}^t)^2-g_{xx}({p}%
^x)^2+i\hbar \theta {\frac{\partial p^\mu }{\partial x^\mu }}=m^2
\end{equation}
which replaces Eqn. (\ref{hjc}) as the dynamical equation.

Here, $g$ is the symmetric part of the metric, and $\theta$ is the value of
a component of $\theta _{\mu \nu}$. The quantum Hamilton-Jacobi equation is 
constructed from here as before, by defining the momentum as gradient of the
complex action. It also appears reasonable to assume that the fundamental
commutation relations (\ref{commu}), as well as the correspondence rule, 
remain unchanged. The new element, on the Planck mass scale, is that the
noncommutative metric now becomes dynamical, and depends on the masses
present. 

We outline here the overall picture for the dynamics of the quantum curved
spacetime, described by the noncommutative metric (\ref{nccm}). This is the
spacetime produced by the particles in the `box Universe' when there is no 
external classical spacetime, and the total mass (and energy) of the 
particles in the box is of the order of the Planck mass scale. This is the
mesoscopic domain, as contrasted to the microscopic (quantum) domain 
considered in the previous section, and as contrasted to the classical 
(macroscopic) domain produced by a system with a mass much larger than the
Planck scale. In analogy with classical general relativity, the metric of the
`quantum gravitational spacetime' (\ref{nccm}) will be determined by the
quantum distribution of particles whose physical state is described by the
complex action $S(\hat{t},\hat{x})$. The evolution of this complex action 
function will in turn depend on the noncommutative metric, as for instance in 
the case of one-particle dynamics described by (\ref{nceq2}) - this
makes the resultant quantum theory of gravity non-linear. The metric and
the dynamical equations are assumed to be covariant under general coordinate
transformations (automorphisms) of the noncommuting coordinates.

A central feature of this dynamics is that the off-diagonal part of the metric,
$\theta_{\mu\nu}$, is assumed to go to zero for large masses. When this 
happens, the quantum dynamics described by Eqns. (\ref{nccm})-(\ref{nceq2})
is indistinguishable from the classical dynamics described using
commuting coordinates on a classical spacetime manifold. Thus we could say
that at the fundamental level dynamics is actually always described using 
noncommuting coordinates, but the use of commuting coordinates in the
macroscopic world is an excellent approximation. It is an excellent 
approximation in the same spirit in which Galilean transformations are
an excellent approximation to Lorentz transformations at speeds much smaller
than the speed of light, or flat spacetime is an excellent approximation to
curved spacetime when the curvature is negligible. 

It is also likely that 
$\theta_{\mu\nu}$ is determined by the imaginary part of the complex matter
action, since the imaginary part of the matter action also vanishes in the 
classical limit. Moreover the presence of the antisymmetric tensor field
$\theta_{\mu\nu}$ for a mesoscopic system implies corrections to the
classical gravitational metric $g$ predicted by general relativity.   

Let us return now to the consideration of the dynamical equation (\ref{nceq3})
and its implications for quantum mechanics. We recall that the complex
action $S(x,t)$ is related to the wave-function by the definition 
$\psi=e^{iS/\hbar}$, and now, because of gravitational corrections which
become important at the Planck scale, the complex action satisfies the new
dynamical equation (\ref{nceq3}) and not the equation (\ref{hjc}). If we  
transform back from (\ref{nceq3}) by substituting for $S$ in terms of $\psi$
we will find that $\psi$ satisfies a non-linear equation, instead of the
linear Klein-Gordon equation (\ref{kg}). This non-linearity is being caused
by the presence of the non-trivial metric components $g$ and $\theta$, and is
in principle observable.

We should now make simplifying assumptions, in order to construct a simple 
example of a non-linear Schrodinger equation. We are
interested in the case $m\sim m_{Pl}$. The symmetric part of the metric - $g$
- should resemble the Schwarzschild metric, and assuming we are not looking
at regions close to the Schwarzschild radius (which is certainly true 
for objects of such masses which we expect to encounter in the
laboratory) we can approximately set $g$ to unity. The key quantity then is 
$\theta =\theta (m/m_{Pl})$ - we have no proof one way or the other whether
$\theta$ should be retained. We will assume here that $\theta$ should be
retained, and work out its consequences. 

$\theta (m/m_{Pl}) $ in principle should also depend on the
quantum state via the complex action $S$, but we can know about the explicit
dependence of $\theta$ on $m$ and $S$ only if we know the dynamical
equations which relate $\theta$ to $m$, which
at present we do not. It is like having to know the analog of the Einstein
equations for $\theta $. But can we extract some useful conclusions just by
retaining $\theta$ and knowing its asymptotic behavior? Retaining $\theta $%
, the above dynamical equation can be written in terms of the complex action 
$S$ as follows: 
\begin{equation}
\label{hjcn}\left( {\frac{\partial {S}}{\partial t}}\right) ^2-\left( {\frac{%
\partial {S}}{\partial x}}\right) ^2-i\hbar \theta (m/m_{Pl}) \left( {\frac{\partial ^2S%
}{\partial t^2}}-{\frac{\partial ^2S}{\partial x^2}}\right) =m^2.
\label{rhj}
\end{equation}
This is the equation we would like to work with. We know that $\theta =1$ is
quantum mechanics, and $\theta =0$ is classical mechanics. We expect $\theta 
$ to decrease from one to zero, as $m$ is increased. It is probably more
natural that $\theta$ continuously decreases from one to zero, as one goes
from quantum mechanics to classical mechanics, instead of abruptly going
from one to zero. In that case we should expect to find experimental
signatures of $\theta $ when it departs from one. By substituting the
definition $S=-i\hbar \ln \psi $ in (\ref{rhj}) we get the following non-linear equation
for the Klein-Gordon wave-function $\psi $: 
\begin{equation}
\label{kg2}-\hbar ^2\left( {\frac{\partial ^2}{\partial t^2}}-{\frac{\partial
^2}{\partial x^2}}\right) \psi +\frac{\hbar ^2}\psi \left( 1-\frac 1\theta
\right) \left( \dot \psi ^2+\psi ^{\prime }{}^2\right) =m^2\psi 
\end{equation}
We are interested in working out the consequences of the non-linearity
induced by $\theta $, even though we do not know the explicit form of $%
\theta $. 

Let us go back to the equation (\ref{hjcn}). In general $\theta $ will also
depend on the state $S$ but for all states $\theta $ tends to zero for large
masses, and if we are looking at large masses we may ignore the dependence
on the state, and take $\theta =\theta (m)$. Let us define an effective Planck's
constant $\hbar _{eff}=\hbar \theta (m),$ i.e. the constant runs with the
mass $m$. Next we define an effective wave-function $\psi
_{eff}=e^{iS/\hbar _{eff}}$ . It is then easy to see from (\ref{rhj}) that the effective wave
function satisfies a linear Klein-Gordon equation
\begin{equation}
\label{psieff}-\hbar _{eff}^2\left( \frac{\partial ^2\psi _{eff}}{\partial
t^2}-\frac{\partial ^2\psi _{eff}}{\partial x^2}\right) =m^2\psi_{eff}
\end{equation}
and is related to the usual wave function $\psi $ through
\begin{equation}
\label{rel}\psi _{eff}=\psi ^{1/\theta (m)}
\end{equation}
For small masses, the effective wave-function approaches the usual wave
function, since $\theta $ goes to unity.

We would now like to obtain the non-relativistic limit for this 
equation. Evidently this limit is
\begin{equation}
ih_{eff}\frac{\partial\psi_{eff}}{\partial t} = 
-\frac{h_{eff}^{2}}{2m}\frac{\partial^{2}\psi_{eff}}{\partial x^{2}}.
\label{psie}
\end{equation} 
By rewriting $\psi_{eff}$ in terms of $\psi$ using the above relation 
we arrive at the following non-linear Schrodinger equation  
\begin{equation}
i\hbar\frac{\partial\psi}{\partial t} = -\frac{\hbar^{2}}{2m}\frac
{\partial^{2}\psi}{\partial x^{2}} + \frac{\hbar^{2}}{2m}(1-\theta)
\left(\frac{\partial^{2}\psi}{\partial x^{2}} - [(\ln\psi)']^{2}\psi
\right).
\label{nlse}
\end{equation}
It is reasonable to propose that if the particle is not free, a term proportional to the potential, $V(q)\psi$, can be added to the above non-linear equation.
In the next section we will get a better insight into the relation between
equations (\ref{psie}) and (\ref{nlse}).

In terms of the complex action function $S$ defined above (\ref{hjc}) as $\psi=e^{iS/\hbar}$ this non-linear Schrodinger equation is written as
\begin{equation}
\frac{\partial S}{\partial t} = - \frac{S'^{2}}{2m} + \frac{i\hbar}{2m}\theta(m)S''\ .
\label{nrhj}
\end{equation}
This equation is to be regarded as the non-relativistic limit of Eqn. (\ref{rhj}).

It is clearly seen that the non-linear Schrodinger equation obtained in this
particular example results from making the Planck constant mass-dependent, in
the quantum mechanical Hamilton-Jacobi equation. Eqn. (\ref{nlse}) is in
principle falsifiable by laboratory tests of quantum mechanics, and its
confirmation or otherwise will serve as a test of the various underlying
assumptions of the noncommutative model. One might be tempted to say at this
stage that the only slender contact that remains with noncommutativity of
spacetime at this stage is the quantity $\theta(m)$ which corrects the
Planck constant in Eqn. (\ref{rhj}) and  non-linearity of the Schrodinger
equation could have been arrived at simply by invoking $\theta$, without
making any reference to noncommutativity. Doing so would of course be ad hoc
and noncommutativity of spacetime provides the underlying reason for the
origin of $\theta$.

Conceptually, it is not difficult to generalize this non-linear Schrodinger
equation to four dimensions, and to the many-particle case. In the following
discussion we will continue to deal with the 2-d non-linear equation
(\ref{nlse}).

\section{A comparison with the Doebner-Goldin equation}

When we found the non-linear equation (\ref{nlse}) we did not know that this
equation already exists in the literature. Only subsequently we learned, from
a review article by Svetlichny \cite{sve}, that many years ago Doebner and 
Goldin arrived at a very similar equation from an apparently different (but 
possibly related) approach. Considering that the same equation has been
arrived at independently by two different routes, and considering that  
the approach of Doebner and Goldin is on firmer ground (as compared to our
partly heuristic analysis) we are led to believe that this non-linear equation
deserves some serious attention, and should be tested in the laboratory.
We will briefly review the Doebner-Goldin equation here, along with its 
possible implications, and its possible connection with our work. Actually
there is an entire class of D-G equations, and we will begin by recalling
the first non-linear equation derived by them.

Doebner and Goldin inferred their equation from a study \cite{dg} of 
representations of non-relativistic current algebras. This involves examining
unitary representations of an infinite-dimensional Lie algebra of vector 
fields $Vect(R^{3})$ and group of diffeomorphisms $Diff(R^{3})$. These
representations provide a way to classify physically distinct quantum systems. 
There is a one-parameter family, labeled by a real constant $D$, of mutually 
inequivalent one-particle representations of the Lie-algebra of probability 
and current densities. The usual one-particle Fock representation is the 
special case $D=0$. The probability density $\rho$ and the current density 
${\bf j}$ satisfy, not the continuity equation, but a Fokker-Planck equation
\begin{equation}
\frac{\partial \rho}{\partial t} = - {\bf \nabla.j} + D\nabla^{2}\rho.
\label{fp}
\end{equation}    

A linear Schrodinger equation cannot be consistent with the above Fokker-Planck equation with $D\ne 0$, but Doebner and Goldin found that the following is
one of the non-linear Schrodinger equations which leads to the above Fokker-Planck equation
\begin{equation}
i\hbar \frac{\partial\psi}{\partial t} = -\frac{\hbar^{2}}{2m}\nabla^{2}\psi + iD\hbar\left(
\nabla^{2}\psi + \frac{|\nabla\psi|^{2}}{|\psi|^{2}}\psi\right).
\label{dg}
\end{equation}

The Doebner-Goldin equation should be compared with the equation (\ref{nlse}) found by us. Although we have considered a 2-d case, and although there are some
differences, the similarity between the two equations is striking, considering
that the two approaches to this non-linear equation are, at least on the face of it, quite different. It remains to be seen as to what is the connection between the representations of $Diff({R^{3}})$, quantum mechanics,  and the antisymmetric part $\theta$ of the asymmetric metric introduced by us. 

The comparison between the two non-linear equations suggests the following relation between the new constants $D$ and $\theta$
\begin{equation}
D\sim \frac{\hbar}{2m}(1-\theta).
\label{dtheta}
\end{equation} 
There is a significant difference of an $i$ factor in the correction terms in
the two equations, and further, the relative sign of the two correction terms 
is different in the two equations, and in the last term we do not have absolute values in the numerator and denominator, unlike in the Doebner-Goldin equation.
Despite these differences, the similarity
between the two equations is noteworthy and we believe this aspect should be
explored further. It is encouraging that there is a strong parallel between the limits $D\rightarrow 0$ and $\theta\rightarrow 1$ - both limits correspond to standard linear quantum mechanics. In their paper Doebner and Goldin note that  the constant $D$ could be different for different particle species. In the
present analysis we clearly see that $\theta(m)$ is labeled by the mass of the particle. 

For comparison with Doebner and Goldin we write down the corrections to the
continuity equation which follow as a consequence of the non-linear terms in
(\ref{nlse}). These can be obtained by first noting, from (\ref{psie}), that
the effective wave-function $\psi_{eff}$ obeys the following continuity 
equation
\begin{equation}
\frac{\partial\ }{\partial t}\left( \psi^{*}_{eff}\psi_{eff}\right)
-\frac{i\hbar_{eff}}{2m}\left(\psi_{eff}^{*}\psi_{eff}' - 
\psi_{eff}\psi_{eff}^{*'}\right)' = 0.
\end{equation}
By substituting $\psi_{eff}=\psi^{1/\theta (m)}$ and $\hbar_{eff}=\hbar\theta(m)$ in this equation we get the following corrections to the continuity 
equation for the probability and current density constructed from $\psi(x)$ 
\begin{equation}
\frac{\partial\ }{\partial t}\left( \psi^{*}\psi\right)
-\frac{i\hbar}{2m}\left(\psi^{*}\psi' - 
\psi\psi^{*'}\right)' = \frac{\hbar (1-\theta)}{m}
|\psi|^{2}\phi''.
\label{cm}
\end{equation}
where $\phi$ is the phase of the wave-function $\psi$, i.e. $\phi=Re(S)/\hbar$.
It is interesting that the phase enters in a significant manner in the
correction to the continuity equation. This equation should be contrasted
with Eqn. (\ref{fp}). The fact that the evolution is not norm-preserving
when the mass becomes comparable to Planck mass suggests that the appropriate  
description should be in terms of the effective wave-function $\psi_{eff}$.

An equation similar to the Doebner-Goldin equation was also independently
derived and studied by Schuch, Chung and Hartmann \cite{sch}, \cite{sch2} 
who suggested
a modification of the continuity equation to include diffusion. They
proposed a non-linear Schrodinger equation with a logarithmic non-linearity
\begin{equation}
i\hbar\partial_t\psi=-\frac{\hbar^{2}}{2m}\nabla^{2}\psi + V\psi
-i\hbar\gamma(\ln\psi-<\ln\psi>)\psi
\end{equation}

\bigskip

\centerline{\bf The general Doebner-Goldin equations}

\bigskip

\noindent The equation (\ref{dg}) is one of an entire class of the non-linear
Doebner-Goldin equations which are consistent with the Fokker-Planck equation
(\ref{fp}). This class is given by \cite{dg2}
\begin{equation}
i\hbar\frac{\partial\psi}{\partial t} = H_{0}\psi + iI[\psi]\psi + R[\psi]\psi
\label{dggen}
\end{equation}
where $H_0$ is the usual linear Hamiltonian operator and $I[\psi]$, $R[\psi]$
are real-valued non-linear functionals given by
\begin{equation} 
R[\psi]=\hbar D'\sum_{j=1}^{5}c_{j}R_{j}[\psi], \qquad 
I[\psi]= \frac{1}{2} \hbar D R_{2}[\psi]
\label{ri}
\end{equation}
where $R_{2}=[\nabla^{2}\rho]/\rho$ and the form of the other terms $R_{i}$
is given in Eqn. (1.5) of \cite{dg2}. The coefficients $D$ and $D'$ are real
numbers with the dimensions of diffusion coefficients. Eqn. (\ref{dg}) is a
special case of the general class given by (\ref{dggen}). It is clear that
our non-linear Schrodinger equation (\ref{nlse}), though very similar to the 
D-G equation, does not belong to the general
Doebner-Goldin class (\ref{dggen}) either, because the corrections to the
continuity equation which our equation implies (Eqn. (\ref{cm})) are not of
the Fokker-Planck type. However, we will see below that there is a 
generalization of the D-G equation (\ref{dggen}) proposed by 
Goldin \cite{gol}, of which our equation (\ref{nlse}) is indeed a special case.

An important concept introduced by Doebner and Goldin, which will be of
relevance to us in understanding the relation between our Eqns. (\ref{nlse})
and (\ref{psie}), is that of a non-linear gauge transformation \cite{dg3}. 
It is important because such gauge transformations relate a set of distinct
non-linear equations as belonging to the same equivalence class, in that
all the equations within a class describe the same physics, with regard to
time evolution. Yet, there are equivalence classes which are inequivalent
to standard quantum mechanics. Another consequence of introducing the
non-linear gauge transformations is that a whole family of non-linear
Schrodinger equations are actually physically equivalent to the standard 
linear quantum mechanics. Also, the coefficients arising in the general
D-G equation (\ref{ri}) are in general not gauge-invariant, and 
gauge-invariant quantities must be constructed from them, before one can
suggest that some of these quantities should be looked for in experiments,
as a test of the proposed non-linearity.    

The construction of the non-linear gauge transformation depends on the
assumption that all quantum mechanical measurements are fundamentally
positional measurements. Thus two quantum theories are considered equivalent
if the corresponding wave functions give the same probability density in
space, at all times. Doebner and Goldin define a two-parameter non-linear
gauge transformation (given by Eqn. (2.2) of \cite{dg3}) which leaves the
probability density invariant. These gauge transformations form a 
two-parameter group, and they construct gauge-invariant parameters out of
the coefficients which appear in their general non-linear equation 
(\ref{ri}).  

\bigskip

\centerline{\bf A generalization of the Doebner-Goldin equation}

\bigskip

Goldin has proposed \cite{gol} a generalization of the D-G equation, by
introducing the terms $R_1$, $R_3$, $R_4$, and $R_5$ of (\ref{ri}) also
in the imaginary part of (\ref{dggen}). The resulting non-linear Schrodinger
equation [Eqn. (10) of \cite{gol}] achieves complete symmetry between the
real and imaginary parts of the non-linear correction terms, and as we will
see shortly, our non-linear equation (\ref{nlse}) is a special case of this
Goldin equation. The Goldin equation is no longer restricted by the 
requirement that it has to be consistent with the Fokker-Planck equation
(\ref{fp}). Thus one is now extending the class of non-linear equations beyond
what was obtained by studying the representations of $Diff{R^{3}}$.
Nonetheless, a suitable generalization of the aforementioned
concept of non-linear gauge transformations allows one to construct a 
conserved probability density which coincides with the standard definition
in the linear theory.

Goldin starts by noting that the amplitude $R$ and the phase $\alpha$ of the
wave function $\psi=Re^{i\alpha}$ are treated asymmetrically in standard
quantum mechanics. $R$ is gauge invariant and physically observable, while
$\alpha$ is not. (We use the notation $\alpha$ for the phase, instead of
Goldin's $S$, as we have already used $S$ to denote the complex action $S$
defined by $\psi=e^{iS/\hbar}$). This asymmetry is unnatural from the point
of view of a non-linear gauge transformation because we are combining a 
gauge-dependent quantity $\alpha$ with a gauge-invariant quantity $R$ to
construct the wave function $\psi$ and through the Schrodinger equation 
coupling both $R$ and $\alpha$ to gauge potentials. It would be more natural
to couple gauge-dependent quantities to each other, and physical quantities
to each other. This, we believe, is an important idea for us, as we have
been dealing with the complex action $S$ and treating its real and imaginary
parts on an equal footing.  

The generalized four-parameter non-linear gauge transformation, after defining
$T=\ln R$ so that $\ln\psi=T+i\alpha$, is given by
\begin{eqnarray}
\left(\begin{array} {cc}\alpha'\\ T'\end{array}\right)= \left(\begin{array}{cc}
                      \Lambda & \gamma \\
                      \lambda & \kappa \end{array} \right)
\left(\begin{array}{cc}\alpha \\ T\end{array}\right)
\label{nlg}
\end{eqnarray}  
The earlier group, which leaves the probability density invariant, is given
by $\lambda\equiv 0$, $\kappa\equiv 1$. In terms of the functions $\alpha$
and $T$, Goldin's non-linear equation can be elegantly rewritten as the
Eqn. (25) in his paper \cite{gol}. It has a complete symmetry between the
amplitude and the phase, or in our language, between the real and imaginary
parts of the complex action $S$. Gauge-invariant quantities can be
constructed from the coefficients in the non-linear equation, and in Goldin's 
notation, the quantities $\tau_1$ and
$\tau_2$ defined by $2\tau_1=a_1 + b_2$ and $2\tau_2=a_1b_2-a_2b_1$ are
gauge invariant. Furthermore, invariant combinations of $\alpha$ and $T$
exist, so that a gauge invariant probability density and a gauge invariant
current can be constructed, and these reduce to the standard definitions
in the linear case.

We can now examine our non-linear equation (\ref{nlse}) as a special case 
of Goldin's Eqn. (25). Our non-linear equation, when expressed in terms
of the complex action $S$, is given by (\ref{nrhj}). By writing the latter
equation in terms of its real and imaginary parts, and comparing with
Goldin's Eqn. (25), we find that our equation is indeed a special case of
Goldin's non-linear equation, with the non-vanishing coefficients given by
\begin{equation}
a_3=-a_5=\hbar/2m, \quad a_2=a_5\theta(m/m_{Pl}), \quad b_4=-\hbar/m,
\quad b_1=-a_2
\end{equation}
 It is easily seen from here that for $\theta=1$ our equation coincides with
the linear Schrodinger equation, as it should.

We now perform Goldin's  non-linear gauge transformation, given in our
(\ref{nlg}) above, with the specific choice 
$\Lambda=\kappa=1/\theta(m/m_{Pl}),\gamma=\lambda=0$. In accordance with
Goldin's Eqns. (28) and (29) this transformation leads to a new evolution
equation with the new non-vanishing coefficients given by
\begin{equation}
a_2'=a_5'=-a_3'=h_{eff}/2m, \qquad b_4'=2b_1'=-h_{eff}/m
\end{equation}
where, as before, $\hbar_{eff}=\hbar\theta(m/m_{Pl})$. Comparison with 
Goldin's (27) shows that the new equation is the linear Schrodinger equation
but with $\hbar$ replaced by $\hbar_{eff}$. This transformation is of course
the same as that in our Eqn. (\ref{rel}) and we now formally understand it as  
a non-linear gauge transformation in the sense of Goldin. A conserved
probability density is hence given, as usual, by $|\psi_{eff}|^{2}$, or
equivalently by $|\psi|^{2/\theta(m)}$.

Using the definitions of $\tau_1$ and $\tau_2$ and the values of
$a's$ and $b's$ given above, the gauge-invariant combinations of 
coefficients are given, in our case, by $\tau_1=0$ and 
$\tau_2=\hbar_{eff}^{2}/8m^{2}$. Thus the quantity $\hbar^{2}/8m^{2}$ by
itself is not gauge invariant. Furthermore, if our ideas are correct,
an experiment designed to measure $\hbar/m$ for a mesoscopic system will
show departure from the standard quantum mechanical value, in the vicinity
of the Planck mass scale. 

It should be noted that the reduction to a linear Schrodinger equation
became possible only for our simplified model, in which $\theta$ depends
only on the mass, without depending on the physical state through the
complex action function $S$. Also, in the more general model described
by Eqn. (\ref{nceq3}) there will be additional non-linear terms introduced
by the symmetric metric components. In general, the non-linear Schrodinger 
equation in the mesoscopic domain will not be equivalent to a linear 
equation via a gauge transformation.  

It is also interesting to ask whether the generalized equation introduced
by Goldin can be arrived at by considerations of symmetry, analogous to
considerations such as representations of $Diff{R^{3}}$ that led to 
the Doebner-Goldin equation. Considering that despite a striking similarity
our non-linear equation differs from the D-G equation in a subtle and   
significant way we would like to conjecture that Goldin's equation may be
related to a study of noncommutative algebras on the manifold, unlike the
commutative algebra that lead to the D-G equation. We hope to investigate
this possibility (which may be paraphrased as a `Hopf generalization of
Lie algebras') in the near future.  

Another interesting question pertains to the field theoretic generalization
of the Doebner-Goldin equation. As we mentioned above, this equation arises
from consideration of inequivalent one-particle representations, which are
related to the representations of the configuration space $R^{3}$. Now $R^{3}$
is also one of the possible background spatial manifolds on which the 3-metric
is defined, in the construction of the Wheeler-DeWitt equation, which of
course is linear. The Wheeler-deWitt equation is then the D=0 analog of
the Doebner-Goldin equation (\ref{dg}), and by considering other 
representations of $Diff{R^{3}}$ one would like to investigate what are the
non-linear generalizations of the Wheeler-DeWitt equation, in the sense
of Doebner and Goldin.

Lastly in this section we would like to briefly remark on some of the earlier
results \cite{nlp}, \cite{pol} that non-linearity can lead to unphysical phenomena such
as 
superluminal propagation. An interesting point is made by Doebner and
Goldin \cite{dg3} when they note that the physical equivalence of some of the
non-linear equations and the linear Schrodinger equation (brought about by
the non-linear gauge transformation) suggests that this argument against
non-linearity cannot be as general as it might have been thought to be. 
More generally, the non-linearity proposed by us arises at the yet untested 
Planck mass/energy scale, and is a consequence of the underlying noncommutative
structure of spacetime. Thus we could not a priori rule out the possibility
that the causal structure of spacetime that we are familiar with may be an
excellent approximation to an underlying quantum spacetime structure which
may in some form permit acausality on the Planck scale.

Some of the other interesting and relevant discussions on
non-linearity in quantum mechanics are those by Bialynicki-Birula and
Mycielski \cite{bia}, Parwani \cite{par} and Adler \cite{adl}.  

\section{The case for experimental tests of mesoscopic quantum mechanics}  

While there are very stringent bounds on non-linearity in the atomic mass range
\cite{wei}, \cite{wei2}, \cite{exp},
it is not really a  well-known fact that quantum mechanics has not been
experimentally verified for objects of intermediate masses, where by
intermediate we mean masses much larger than atomic or molecular masses, and
much smaller than macroscopic masses. This range indeed spans many orders
of magnitude, and an interesting range to explore would be say from
$10^{-15}$ grams to $10^{-8}$ grams. Most people do not expect any departures
from quantum mechanics in this domain, but should we pre-judge the results
of experiments in a range of parameters that has not been explored, and for
which there are now theoretical reasons for further investigation? There 
seems to be a general impression that enough is known about systems such as
macromolecules, polymers, nanoparticles etc. so that any deviations from
quantum mechanics would have been seen by now. However it should be 
emphasized that none of these systems fall in the requisite mass range, 
despite an impression to the contrary. Thus, if we could make an object of    
say a billionth of a gram, isolate it from its environment, and arrange things
so that its internal degrees of freedom can be ignored (so that it behaves
like a point particle), we do not know from any experiment whether or not its 
dynamics obeys the rules of standard quantum mechanics.  

In Section 2 we have argued that, because of gravitational effects, quantum 
mechanics becomes non-linear at the Planck mass/energy scale. This argument
is generic, and does not depend on the particular noncommutative model we have
described subsequently. The genericity of our arguments suggests to us that
there is a strong case for an experimental investigation; at the very least
one will succeed in putting bounds on non-linearity in the mesoscopic
domain. In the context of our specific model described in Section 5 we have
predicted that Planck's constant runs with mass (equivalently, energy) and
$\hbar\theta(m/m_{Pl})/m$ is a gauge invariant quantity which can in principle
be measured experimentally. 

Another general impression is that since mesoscopic objects have physical
sizes much, much larger than their associated de Broglie or Compton 
wavelengths, quantum effects, or departures therefrom, will be essentially
impossible to detect. This is true to a large degree, but not entirely.
Experiments, such as an interference experiment, which rely on having
an experimental set-up with sizes comparable to the wavelength of the particle
wave will indeed not be of help. But there are other possible measurements,
such as the energy spectrum, or particle-particle scattering, where physical
size need not play a direct role, and which could carry information about
the non-linearity. Through future work we hope to be able to make predictions 
of more gauge-invariant quantities which carry information about the
non-linearity and can be subjected to experiment. In the meanwhile, we feel
there is already a case for experimentally investigating if (and how) one
can design a mesoscopic object which behaves like a point-particle (i.e.
internal degrees of freedom can be ignored) and which can be isolated 
from its environment. It seems likely that such objects could be designed
more easily by smashing macroscopic grains (as opposed to congregating
atoms).      

The antisymmetric tensor field $\theta_{\mu\nu}$ introduced by us, which is 
responsible for the quantum non-linearity, becomes weaker with increasing 
mass, unlike the usual gravitational field described by the metric 
$g_{\mu\nu}$. This antisymmetric field hence violates the equivalence
principle, and search for such a field in the mesoscopic domain is another 
possible test for our ideas. The presence of a field of this nature also
indicates there will be a departure from the inverse square law of
gravitation of an object, as the value of the mass is brought closer to the
mesoscopic range. 

A remark on quantum measurement and wave-vector reduction. Penrose has 
suggested \cite{pen} that gravity may be responsible for the collapse of the 
wave function during a quantum measurement. While it is not clear to us that
the non-linearity in our non-linear equation plays a role during a 
quantum measurement, it seems that such an equation, in which Planck mass
scale non-linearities are induced as a result of self-gravity, may provide
a suitable mathematical set-up for theoretically implementing Penrose's 
idea.     

\section{Concluding Remarks}

In this paper we have argued that there should exist a formulation of
quantum mechanics which does not refer to a classical spacetime manifold. 
Physicists are likely to generally agree with this claim. Disagreement may
arise with our subsequent inference that this implies quantum gravity to
be a non-linear theory. It could be said that the non-linearity discussed here
arises as a result of the gravitation self-interaction, and is not intrinsic 
to quantum mechanics. In a way that is right, but then mass-energy and
gravitation are 
universal features of any physical system (classical or quantum), and we are
saying that the dynamics on the Planck scale is a generalization of
standard linear quantum dynamics, to which the linear dynamics is an
excellent approximation at energy scales much smaller than the Planck scale.
Perhaps the relation between the Planck scale dynamics and the linear quantum
dynamics is similar to the relation between special relativity and general 
relativity. The geodesic equation of motion in the presence of a gravitational
field is of course different from the corresponding equation in the absence of
a gravitational field, though the former can certainly be called a 
generalization of the latter. Similarly, quantum dynamics in the presence of
a quantum gravitational field is different from quantum dynamics in the
absence of such a field. Furthermore, the linear dynamics and the non-linear 
dynamics will predict different results for the same experiment.    

Despite appearances, our emphasis is not overly and exclusively on the
non-linearity, but in trying to apply noncommutative differential
geometry to obtain a new formulation of quantum mechanics. Non-linearity is then a byproduct.
Undoubtedly, our ideas on the use of noncommutative geometry in this context
are still largely heuristic, and remain to be put on a proper mathematical
foundation. Nonetheless, the physical content of these ideas appears
attractive and persuasive. These ideas generalize special relativity and
general relativity in much the same way that noncommutative differential
geometry generalizes Riemannean geometry. That there is a connection with the
Doebner-Goldin equation, which was derived completely independently and with
a different motivation, is encouraging. An important task is to find out the
field equations analogous to the Einstein equations, which the asymmetric 
tensor $h_{\mu\nu}$ should satisfy - equations that should reduce to Einstein 
equations in the classical limit.

Finally, a few remarks comparing our ideas with two of the much better 
developed approaches to quantum gravity - loop quantum gravity and string
theory - which are both based on a linear quantum theory. LQG investigates
whether standard quantum theory and general relativity can together be
merged into a consistent theory of quantum general relativity - undoubtedly
a well-defined question. It appears to us that from the point of view of
the Doebner-Goldin equation, LQG is the D=0 limit of an entire class of
inequivalent quantum gravity theories. It would be of major interest to ask
how the linear theory would relate to the non-linear theory which would
result from considering the feedback of the quantum gravitational field
on itself, in the sense discussed in this article.

In the context of string theory, it is interesting that the Doebner-Goldin 
equation arises as an effective non-linear Schrodinger equation when one 
considers
the quantum dynamics of a system of D0-branes \cite{mav}. An effective 
Fokker-Planck equation for the probability density is arrived at when one
considers the quantum recoil due to the exchange of string states between
the individual D-particles. To us the spirit here seems to be very much
the same as that of our earlier question: how do quantum gravitational
effects modify the quantum dynamics of the particles in our box Universe.
That the Doebner-Goldin equation arises in both the analyses raises an
interesting question - is one discussing the same physical situation in two
different languages? Indeed, since one expects a new picture of a quantum
spacetime to emerge from string theory, one could ask if noncommutative
geometry is a natural language for an ab initio description of the quantum 
spacetime, from which classical spacetime emerges in an approximation. This
would be a top-down approach (as advocated in this article) as opposed to
the more conventional, and better understood, bottom-up approaches in which one
starts with a linear quantum theory, and then allows for the possibility that
feedback effects can introduce non-linearity. (An elementary example comparing
top-down and bottom-up approaches could be had from general relativity. One
can study non-linear gravitational effects by successive iterations of
a linearized theory of gravitation, or one can obtain linearized gravity 
as an approximation to the full theory. Obviously, the physical principles
that go into the construction of general relativity are independent of our
theoretical knowledge of linearized gravity).
A possibility of a 
modification of quantum mechanics in string theory has been considered also
in \cite{ell} as resulting from coupling to massive string states in the
`spacetime foam'. 

The work described in this article is partly based on
a few earlier preliminary papers \cite{qm1}, \cite{qm2}, \cite{qm3}, 
\cite {qm4} and was partly carried out in collaboration
with Sashideep Gutti and Rakesh Tibrewalla \cite{qm5}.

\bigskip

\noindent{\bf Acknowledgements:} It is a pleasure to thank Vladimir Dobrev,
Hanz-Dietrich Doebner and other organizers of the Varna Conference for 
hospitality, and for organizing a stimulating conference. For useful
discussions at the meeting I would like to thank Hanz-Dietrich Doebner,
Klaus Fredenhagen, Gerald Goldin, Harald Grosse, Dieter Schuch and Tony
Sudbery. For
discussions, conversations or correspondence on various earlier occasions I 
thank Abhay Ashtekar, Sergio Doplicher, Sashideep Gutti, Claus Kiefer, 
N. Kumar, Vandana Nanal, T. Padmanabhan, Rajesh Parwani, R. G. Pillay, 
Joseph Samuel, George Svetlichny, R. Srikanth, Sumati Surya, Rakesh 
Tibrewalla, C. S. Unnikrishnan, K. P. Yogendran and Cenalo Vaz. It
should be added that not everyone thanked here necessarily agrees with
the line of thought developed in this article!

\bigskip

\noindent{\bf A note on the references:} The literature on physical 
applications of noncommutative geometry is vast indeed, and there are
also a considerable number of papers on non-linear quantum mechanics. The
list of references below is certainly far from being complete, and is only
suggestive of a few of the related ideas existing in the literature.

The statement of Einstein cited at the beginning of this article is from
his paper in {\it J. Franklin Inst.} 221, 313 (1936). It is mentioned also 
in the book `Subtle is the Lord' by Abraham Pais (p. 461).

\end{document}